\documentclass[aps,prb,twocolumn,superscriptaddress,floatfix]{revtex4-1}
\usepackage{amsmath,amssymb,graphicx,graphics,color}
\usepackage{dcolumn}
\usepackage{bm}
\usepackage{psfrag}

\begin{document}

\title{Topological quasi-one-dimensional state of interacting spinless electrons}

\author{G. Sun}
\affiliation{Max-Planck-Institut f\"ur Physik komplexer Systeme, Dresden, Germany}

 \author {T. Vekua}
\affiliation{Institut f\"ur Theoretische Physik, Leibniz Universit\"at Hannover, Germany}

\begin{abstract}
By decreasing the transversal confinement potential in interacting one-dimensional spinless electrons and populating the second energetically lowest sub-band, for not too strong interactions system transitions into a quasi-one-dimensional state with dominant superconducting correlations and one gapless mode. By combining effective field theory approach and numerical density matrix renormalization group simulations we show that this quasi-one-dimensional state is a topological state that hosts zero-energy edge modes. We also study the single-particle correlations across the interface between this quasi-one-dimensional and single-channel states.
\end{abstract}

\maketitle

\date{\today}


\section{Introduction}

 One surprising result of the integer quantum Hall effect studies is that band insulators of electrons with filled Landau levels can possess non-trivial topological properties \cite{QHE}, which in particular make quantum Hall systems important tools in metrology. 

After the work of Haldane \cite{Haldane}, where it was shown that Landau levels are not a necessary ingredient for the non-trivial topological nature of band insulators, many more electron systems have been shown to possess topological properties, including those which do not rely on broken time-reversal invariance and can occur as in bulk insulators \cite{KF,BZ,Koenig,HK}, so in superconductors with fully gapped \cite{Kitaev} or partially gapped spectrum \cite{Fid,Sau,ChengTu} within the particle-number-non-conserving (BCS-type) as well as particle-number-conserving models. The most common property of these topological states is presence of the gapless degrees of freedom associated with the boundaries. The emergence of the new state of the matter, topological state, apart of the deeper understanding of the behaviur of many-electron systems, can also find application in, e.g., quantum computation due to topologically protected Majorana quasiparticles that they can host \cite{Sarma}.

In this work, we revisit the problem of the phase transition between the single-channel and quasi-one-dimensional states of interacting quantum wire of spinless electrons with decreasing the transversal confinement frequency \cite{BalFish,Starykh,Ledermann,Balents,Meyer,Rosch,Meyer1,Meyer2} in the regime when interactions are not too strong and two-band description can be applied. Using combination of effective field theory bosonization and density matrix renormalization group (DMRG) method \cite{White,Uli}, we show that when starting populating the second lowest transverse sub-band, the quasi-one-dimensional (q1D) ground state with single gapless mode (a superconductor with quasi-long-range order in pairing field) is (or at least is connected adiabatically, without phase transition to) a topological state (Tq1D) characterised by the double degeneracy and emergence of the zero-energy edge modes for open boundary conditions. In contrast, for periodic boundary conditions the double degeneracy of the ground state is lifted. We also check that the above physics is stable with respect to weak time-reversal symmetry-breaking perturbations and characterize numerically quantum phase transition from single-channel regime to Tq1D state. At the end of our work we study the single-particle correlation functions for the case when a spatial interface between the single-channel and Tq1D regimes is present.

\begin{figure}
\includegraphics[width=7cm]{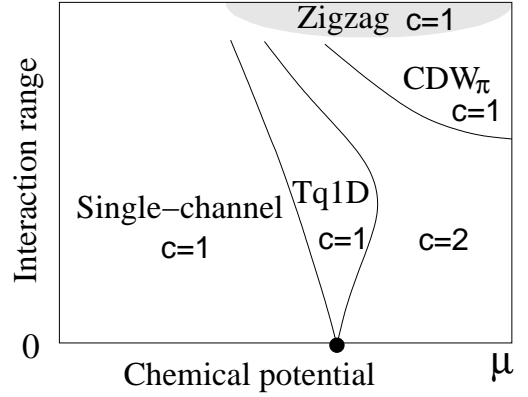}
\caption{A sketch of the phase diagram of interacting spinless electrons as function of interaction range and chemical potential or transversal confinement length in the regime when interactions are not too strong and restriction to the two-band model can be justified. Tq1D stands for quasi-one-dimensional superconducting state and CDW$_{\pi}$ stands for out of phase charge-density wave. In each phase, central charge $c$ indicates the number of gapless modes present. The zigzag state indicated by the gray area is realized in the regime of strong interactions and is outside of the two-band approximation.}
\label{fig:pd}
\end{figure}

In experiments on quantum wires one can change chemical potential of electron gas by applying a gate voltage. For strong Coulomb interactions and small chemical potential, electron gas forms a crystal like one-dimensional structure for strong transversal confinement: the so-called Wigner crystal \cite{Piacente} melted by quantum fluctuations at large distances \cite{Schulz}. With increasing the electron density, or equivalently decreasing the transversal confinement frequency, electron gas in the regime of strong interactions explores the second dimensionality and enters the quasi-one-dimensional planar state with developing a zigzag structure \cite{Piacente}. The signatures of such transitions were revealed experimentally in low-temperature transport properties of one-dimensional quantum wires, in
GaAs/AlGaAs heterostructures, as the confinement strength and the carrier density were varied \cite{exp1,exp2}. Both the linear and zigzag phases have only one gapless excitation mode, which corresponds to longitudinal sliding of the crystal. The classical transition between the linear and zigzag structures, induced by decreasing the transversal confinement frequency, was explored in  experiments on ionic chains \cite{Birkl,Enzer,Mehl}. For the dipolar systems, using Monte Carlo simulations, it was observed that quantum fluctuations smoothen the classical transition between linear to planar structures \cite{Astra}. Approximating transverse direction with discrete (up or down) variable, using two leg ladder lattice, it was suggested that quantum fluctuations melt zigzag like order of ladder site densities, however, string order emerges \cite{Ruhman} as reminiscent of transversal ordering in the ``particle basis''. Using Monte Carlo simulations, it was shown that long-range order of transverse coordinates of electrons indeed survives quantum fluctuations present in the zigzag phase \cite{Meyer2}. In Fig. \ref{fig:pd}, we sketch ground-state phases of spinless electrons as function of interaction range and chemical potential based on the previous studies \cite{BalFish,Starykh}. In our numerical simulations, restricted to two-band lattice model, we observe all the phases depicted in Fig. \ref{fig:pd}, except of the zigzag planar phase, for which it is crucial to include second dimension and hence it is outside of the description based on purely one-dimensional two-band model.

\section{Universal effective theory  describing transition from single-channel regime to q1d state for not too strong interactions}

In the regime when interactions are small compared to Fermi energy, to describe the vicinity of the quantum phase transition between single-channel and q1D states, instead of developing a zigzag structure in the second dimension in the q1D state, the appropriate picture is to think of the starting filling the second lowest transverse sub-band with increasing the chemical potential. The properties of electron gas, apart of the density and band structure of the two sub-bands, are described by interaction constants accounting for density-density interactions in the lowest, second and between the two sub-bands $\sim g_x$ and a pair tunneling $\sim \gamma_t$ between the two sub-bands. The universal effective theory for electrons interacting with screened Coulomb repulsion, describing the vicinity of the quantum phase transition between the one-dimensional and q1D states, is given by the following Bose-Fermi Hamiltonian \cite{Meyer}:

\begin{eqnarray}
\label{BosFer}
{\cal H}&=&\frac{\hbar v_{F1}}{2} \int \mathrm{d}x \left( K_1 (\partial \theta_1)^2+\frac{(\partial \phi_1)^2}{K_1} \right)\\
&+& \int \mathrm{d}x \psi_2^{\dagger} ( - \frac{\hbar^2 \partial^2 }{2m}-\mu+\mu_{c} ) \psi_2\nonumber\\
&+& \gamma_t \int \mathrm{d}x \left[ (  \partial \psi_2 \psi_2- \psi_2 \partial \psi_2) e^{2i\kappa \theta_1}+\mathrm{H.c.}\right]. \nonumber
\end{eqnarray}

Indices indicate sub-bands of transversal confinement. First (energetically lowest) sub-band is always partially filled and is described by bosonic fields ($\theta_{1},\phi_{1}$) corresponding to phase and density fluctuations in the lowest sub-band, respectively, with commutation relations $[\theta_{1}(x),\partial_y\phi_{1}(y)]=i\delta(x-y)$. The Luttinger liquid constant $K_1$ as well as sound velocity $v_{F1}$ are in general complicated functions of microscopic parameters that should be fixed by comparing with numerics. The second sub-band, while crossing the phase transition, in the absence of interaction experiences vacuum to finite density commensurate-incommensurate (C-IC) phase transition (indicated by filled bullet in Fig. \ref{fig:pd}) and is adequately described by the fermionic degrees. The parameter $\kappa$ is related to the strength of density-density interaction between the sub-bands $g_x\sim 1-\kappa$.

For $\mu<\mu_c$, at low energies, the second transverse quantization mode is negligible and ground state is a trivial phase corresponding to a Luttinger liquid 
of interacting single-component fermions. However, when  $\mu>\mu_c$, the second transverse quantization mode can not be ignored any more from the low-energy description.
Despite this, bosonization analyses \cite{BalFish,Meyer} reveal that in the presence of interactions on both sides of phase transition, the system is described by a one-component Luttinger liquid with one gapless and one gapped mode. Density-density type interactions can not open a gap, hence, a crucial term at low energies is the pair-hopping term $\sim \gamma_t$ that opens a gap in the second mode for $\mu>\mu_c$, where critical value $\mu_c$ gets renormalized by interactions. One of the aims of our work is to characterize numerically the quantum phase transition between two single-component Luttinger liquid states: purely one dimensional and q1D. Our main goal is to establish the topological nature of the latter state.

\section{Lattice Hamiltonian for numerical simulations}

\begin{figure}
\includegraphics[width=7.50cm]{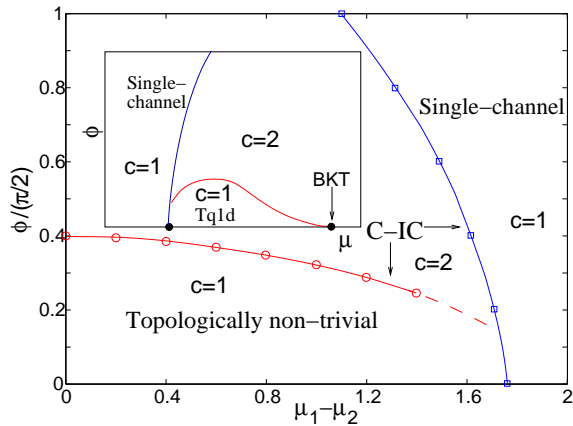}
\caption{ Effect of the time-reversal symmetry breaking on the transition from single-channel to two-channel regimes. Microscopic parameters are $W=1.8$, $g_x=0$ and filling is $1/3$ in the minimal lattice model Eq. (\ref{eq:hamiltonian}). For $\phi=0$ there is a direct transition between two $c=1$ one-component Luttinger liquids. Dashed line is guide for eyes in the region where our numerical procedure becomes less reliable. Inset: the phase diagram expected when including density-density type interactions. Two phase transitions for $\phi=0$, indicated by filled bullets, merge into a single C-IC point for non-interacting electrons indicated in Fig. \ref{fig:pd}. Similar phase diagram is obtained when changing $W$ and filling.}
\label{fig:model1}
\end{figure}

The single-band lattice model describing interacting one-dimensional spinless electrons is given by the following Hamiltonian,
\begin{equation}
H=-t\sum^L_{r=1}(  a^\dagger_{r+1} a_{r}+H.c.)+  \sum^L_{r=1,\tilde r=1,2,...}V_{\tilde r} a^\dagger_{r}a^\dagger_{r+\tilde r}a_{r+\tilde r}a_{r},
\end{equation}
where $a_r$ and $a^\dagger_{r}$ are fermionic annihilation and creation operators at site $r$, $t$ is the hopping rate, and $V_{\tilde r}$ is interaction potential.
In the presence of two relevant sub-bands, we decompose $a_r\to w_1 a_{1,r}+w_2 a_{2,r}$, where $a_{1,r}$ and $a_{2,r}$ are operators corresponding to two lowest transversal quantization sub-bands, and coefficients $w_1$ and $w_2$ are such that single-particle processes are diagonal in the sub-bands basis,
\begin{eqnarray}
\label{exact}
H=&-&\sum^L_{\ell=1,2;r=1}(t_{\ell}  a^\dagger_{\ell, r+1} a_{\ell, r}+\mathrm{H.c.}) \nonumber\\
 &+&g_x \sum_{r=1}^L n_{1,r} n_{2 ,r} -\sum_{\ell=1,2;r=1}^L \mu_{\ell}n_{\ell ,r}\nonumber\\
&+& \!\!\!\!\!\!\!\!\! \sum^L_{\ell_j=1,2;r=1,\tilde r=1,2,...} \!\!\! \!\!\!\!\! V^{\ell_1,\ell_2,\ell_3,\ell_4}_{\tilde r}  a^\dagger_{\ell_1,r}a^\dagger_{\ell_2, r+\tilde r}a_{\ell_3,r+\tilde r}a_{\ell_4,r}, 
\end{eqnarray}
where $n_{\ell,r}=  a^\dagger_{\ell,r} a_{\ell,r}$ are local particle densities and $\mu_l$ are chemical potentials controlling occupations of the sub-bands. Our aim is to study the low energy properties of the ground state of the system when second quantization sub-band starts to get populated, properties that are accounted by the universal effective field theory Eq. (\ref{BosFer}). Let us discuss the interaction part of Hamiltonian (\ref{exact}). We can identify among interaction terms processes, which similar to the kinetic energy, conserve the relative parity symmetry of the two sub-bands. These are processes which do not involve transfer of odd number of fermions between the two sub-bands. However, there are also processes which are called assisted inter-band tunneling with amplitudes $V^{1,1,1,2}_1, V^{1,1,2,1}_1, V^{1,2,2,2}_1, V^{2,1,2,2}_1$ (plus those obtained from exchanging indices $1\leftrightarrow 2$ in the above amplitudes), which involve transfer of odd number of fermions between the sub-bands and hence do not preserve the relative parity symmetry of the two sub-bands. To investigate low-energy properties of the model (\ref{exact}), we will make drastic simplifications and only retain terms that are responsible for determining the infrared thermodynamic properties: these are density-density type interactions and pair-hopping processes \cite{Meyer,Rosch}. Moreover, we will assume the extreme case of very strong screening. After determining the ground-state properties for the simplified model (where we ignore relative parity breaking terms), we will study their effects on the nature of the ground state phases. In addition, the boundaries of the open chain can induce scattering from one sub-band to the other. We will also address the effect of the single-particle tunneling at the boundaries.

For the lattice Hamiltonian that we will use in our numerical simulations we will take a minimal model that captures both the phase transition (between single-channel and q1D states) as well as the single-component Luttinger-liquid ground state phases in its vicinity: 
\begin{eqnarray}
\label{eq:hamiltonian}
H &=& -J  \sum_{\ell=1,2;r=1}^L \left( e^{i(-1)^{\ell} \phi/2 } a^\dagger_{\ell,r+1} a_{\ell,r} + \mathrm{H.c.} \right) \nonumber \\
   && +\sum_{r=1}^L \left( W a^\dagger_{1,r}  a^\dagger_{1,r+1}   a_{2,r+1}   a_{2,r}  + \mathrm{H.c.} \right)  \\
&&+g_x \sum_{r=1}^L n_{1,r} n_{2 ,r} +\sum_{l,l',r}g_{l,l'}n_{l,r}n_{l',r+1}-\sum_{\ell=1,2;r=1}^L \mu_{\ell}n_{\ell ,r}.\nonumber
\end{eqnarray}
The geometry of the lattice is a two-leg ladder with $L$ rungs and $a^\dagger_{\ell,r}$ creates a fermion on site $\ell=1,2$ of the $r$-th rung. To study effects of time-reversal 
symmetry breaking we have introduced nonzero flux $\phi$ per ladder plaquette. We assumed for simplicity $t_{1}=t_{2}=J$ and introduced the denotations $W\sim V_1^{1,1,2,2}=V_1^{2,2,1,1} \sim \gamma_t$, $g_{l,l'}\sim  V_1^{l,l',l',l}$ for the nearest-neighbour pair-tunneling and interactions respectively. The ladder legs play the role of the transversal quantization sub-bands with lower leg corresponding to the lowest sub-band and upper leg to the second sub-band. The relative chemical potential $\mu_1-\mu_2$ controls the imbalance of the particle densities on the ladder legs. Throughout the paper, we will set $J=1$.

Applying the bosonization procedure to the fermions on the lower leg of the ladder in the simplified lattice model Eq. (\ref{eq:hamiltonian}), while keeping the fermionic description for the particles on the upper leg and taking the continuum limit gives effective theory Hamiltonian Eq. (\ref{BosFer}) in the case of the vanishingly small particle density in the upper leg. One and the same effective theory  Eq. (\ref{BosFer}) corresponds to our minimal lattice model Eq. (\ref{eq:hamiltonian}) and the microscopic model of interacting two-band fermions Eq. (\ref{exact}) for the case when particle density in the second sub-band is vanishingly small. Due to this reason, in order to study the universal low-energy properties of the ground state phases and phase transition between them described by the effective theory  Eq. (\ref{BosFer}) we will use the simplified minimal model (\ref{eq:hamiltonian}). After establishing the nature of the ground states with the help of the simplified lattice model, we will address the effects of the terms that we have omitted in Eq. (\ref{eq:hamiltonian}) later on, terms that do not modify the low energy thermodynamic properties, but can influence the topological properties, since they do not preserve the relative parity symmetry.

First, we discuss the $g_{x}=g_{l,l'}=0$ case, as pair-hopping term is the most relevant one, responsible for opening of the excitation gap in the vicinity of the quantum phase transition when the second sub-band starts to fill. The ground-state phase diagram in the parameter plane of $\phi$ and $\mu_1-\mu_2$ is presented in Fig. \ref{fig:model1}, where we identify three different phases, each characterized by the number of gapless modes counted by the central charge $c$ that we have calculated numerically (data not shown).

The model (\ref{eq:hamiltonian}) for  $g_{x}=g_{l,l'}=0$, $\phi=0$ and $\mu_1-\mu_2=0$ was argued to be realizing a topological state with localized zero-energy edge modes (interpreted as Majorana quasiparticles) at the boundary and doubly degenerate entanglement spectrum \cite{Dalmonte} away of half-filling, whereas the appearance of edge modes in the continuous limit was predicted within bosonization \cite{ChengTu}. It was shown that the topological state was stable with respect to weakly disordering local chemical potentials and weak uniform single-particle tunnelings between the legs. We confirm the above properties in our simulation and further show that in the absence of single-particle tunneling between the channels, the topological state is stable with respect to weak time-reversal symmetry breaking and, most importantly, with varying $\mu_1-\mu_2$ it extends all the way until transitioning into the single-channel state for $\phi=0$. The relevant symmetry protecting the topological state is the $Z_2$ relative particle parity symmetry \cite{Dalmonte,ChengTu}. Recently exactly solvable number conserving two-wire fermionic models have been presented, featuring the same properties \cite{Lang,Iemini} and allowing fully analytic treatment, in  particular, identification of Majorana-like quasiparticles.

\section{Bosonization approach for the finite filling of the second band}

When the filling of the second sub-band becomes finite the convenient basis for effective field theory formulation in this case is provided by symmetric and antisymmetric combination of the bosonic fields describing lowest and second transverse quantization modes. 
Introducing another pair of conjugate bosonic fields ($\theta_{2},\phi_{2}$), describing phase and density fluctuations of fermions in the second sub-band, the low-energy properties of the model Eq.~(\ref{eq:hamiltonian}) are then governed by the following Hamiltonian density,
\begin{eqnarray}
\label{EFF}
{\cal H}&=&\frac{\hbar v_{+}}{2\pi} \int \mathrm{d}x \left( K_+ (\partial \theta_+)^2+\frac{(\partial \phi_+)^2}{K_+} \right)\nonumber\\
&+&  \frac{\hbar v_{-}}{2\pi} \int \mathrm{d}x \left( K_- (\partial \theta_--\frac{\sqrt{2}\phi}{\sqrt{\pi}})^2+\frac{(\partial \phi_-)^2}{K_-} \right)\nonumber\\ 
&+&\alpha \gamma_t \int \mathrm{d}x \cos{\sqrt{8\pi} \theta_-},
\end{eqnarray}
where  $\phi_\pm=(\phi_1\pm \phi_2)/\sqrt{2}$, $\theta_+=(\theta_1+ \theta_2)/\sqrt{2}$, $\theta_-=(\theta_1- \theta_2 +\phi x/\sqrt{\pi})/\sqrt{2}$, and $\alpha$ is a numerical normalization constant.
$K_{\pm}$ are Luttinger-liquid parameters corresponding to the total and relative fluctuations of the two-leg ladder and $v_{\pm}$ are corresponding velocities.

In deriving this expression, it was assumed that Fermi wave vectors of the two transverse quantization modes are strongly mismatched $k_{F1}\neq k_{F2}$, as fillings of the two sub-bands are completely different, and there are no commensurability effects with underlying lattice. In case when $k_{F1}\simeq k_{F2}$, for long-range interactions and small values of interaction amplitude, instead of superconducting pairing, an out-of-phase charge-density-wave CDW$_{\pi}$ between the two legs of the ladder dominates \cite{Starykh} due to the competing contribution in the antisymmetric sector $\sim g_x \cos{(\sqrt{8\pi} \phi_-+2(k_{F1}- k_{F2})x)}$. The phase transition from the Tq1d state to CDW$_{\pi}$ state can be induced by increasing amplitudes of $g_x$ and $g_{l,l'}$ in the lattice Hamiltonian Eq. (\ref{eq:hamiltonian}) relative to the pair-hopping amplitude $\gamma_t$ for $\mu_1\simeq \mu_2$. CDW$_{\pi}$ phase is also characterized by one gapless mode and has a long-range non-local bosonic string order \cite{Ruhman} $|\langle (n_{1,i}-n_{2,i})e^{i\pi \sum^j_{l=i}(n_{1,l}+n_{2,l})}  (n_{1,j}-n_{2,j})\rangle|$. 
Like in the Tq1d state, in CDW$_{\pi}$ phase, the single fermion correlation function decays exponentially with distance, however, as opposed to the behavior in the Tq1D state, there is no superconducting quasi-long-range ordering. Moreover, the properties of CDW$_{\pi}$ phase do not depend crucially on statistics, unlike the behavior in the Tq1D state.

As mentioned above, for interacting spinless electrons in quantum wire, for arbitrary non-zero interaction strength, the vicinity of the quantum phase transition point $\mu=\mu_c$ from both sides is described by a single-component Luttinger liquid state \cite{Meyer}. In the single-component Luttinger liquid of the q1D state, the symmetric mode is gapless, while antisymmetric mode is gapped, namely, the relative phase fluctuations of the two sub-bands are locked in the ground state, $\langle \theta_-\rangle =0$ or $\pi$ depending on the sign of $\alpha \gamma_t$. 

The Tq1D superconducting phase is characterized by emergence of non-local string order
\begin{equation}
\label{stringorderfermions}
O^r_s=\langle   (a^\dagger_{1,i} a_{2,i}+\mathrm{H.c.}) e^{i\pi\sum^{i+r}_{j=i}(n_{1,j}+n_{2,j})} (a^\dagger_{1,i+r} a_{2,i+r}+\mathrm{H.c.}) \rangle.
\end{equation}

Note that the similar effective bosonization formulation Eq. ({\ref{EFF}}) is applicable to hard-core bosons in the two-channel regime, however, the latter case is not characterized by topological properties, instead the bosonic q1D state shows twofold-degenerate ground state as for open so for periodic boundary conditions and no localized edge modes. Moreover, the single boson correlation functions are not gapped out, as opposed to fermions \cite{Berg,ChengTu}. In the case of bosons one can define a local order parameter that picks up nonzero value in superconducting q1d state, e.g. for $\alpha\gamma_t<0$ 
\begin{equation}
\langle   b^\dagger_{1,r} b_{2,r}+\mathrm{H.c.}\rangle=\pm \kappa,
\label{pm}
\end{equation}
where $\kappa \sim \langle  \cos{\sqrt{2\pi} \theta_-} \rangle\neq 0$ and $+$ sign in Eq. ({\ref{pm}}) corresponds to one of the ground states and $-$ to another. 

From the expression of the order-parameter equation (\ref{pm}), it is clear that for bosons the uniform single-particle tunneling along the rung will lift the two-fold degeneracy of the ground states immediately, similar to the effect of magnetic field applied to the classical Ising chain with two-fold degenerate ground state. For fermions, as already mentioned, the double degeneracy of the ground state in the Tq1D state (which only holds for the open boundary conditions) is immune against weak uniform single-fermion tunneling along the rung \cite{Dalmonte,Chen}.

Most importantly, the bosonization expression of our model (\ref{EFF}) shows that for small values of fluxes, the Tq1D state is stable with respect to changing interaction strengths even deep in the phase $\mu>\mu_c$ as long as $K_-\ge 1$. On the other hand, using Eq. (\ref{BosFer}) for zero flux case, it has been shown \cite{Meyer} that this same Tq1D phase adiabatically extends towards the vicinity of quantum critical point for arbitrary (non-zero) values of interactions. The effect of flux is that after $\phi>\phi_c$, the two-component Luttinger-liquid state ($c=2$) is stabilized via the C-IC phase transition from the topological state.

\section{Numerical results}

The numerical data that we will present are obtained for minimal lattice model (\ref{eq:hamiltonian}) for large $W$ and in most cases $g_x=g_{\ell,\ell'}= 0$ (unless stated otherwise) since topological features are most pronounced for this case. We have included also in simulations the density-density type interactions $g_x$ and $g_{\ell,\ell'}\neq 0$ in Eq. (\ref{eq:hamiltonian}) and checked that similar properties of the ground state remain stable, although become less pronounced with increasing their strengths (until system transitions into the two-component Luttinger-liquid state or CDW$_{\pi}$ state for $k_{F1}\simeq k_{F2}$ for strong density-density interactions, where topological features disappear).

\begin{figure}
\includegraphics[width=8.0cm]{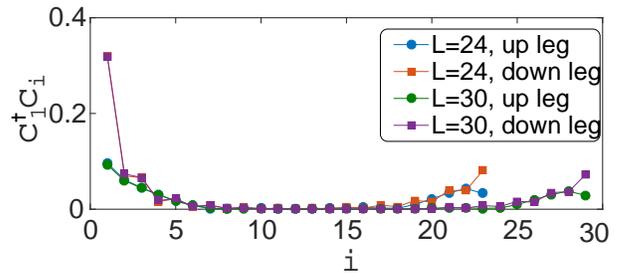}
\caption{Subband-resolved single-particle correlation functions starting from an edge for $\mu_1-\mu_2=0.8$ and $\phi=0.1\pi/2$ for $W=1.8$ in Eq. (\ref{eq:hamiltonian}). Edge-to-edge correlations show recovery, due to the edge modes.}
\label{fig:edgetoedge}
\end{figure}

In Fig. \ref{fig:edgetoedge}, we present single-particle correlation functions obtained for Eq. (\ref{eq:hamiltonian}) for $\mu_1-\mu_2=0.8$ and small value of the flux, starting from the edge. We see that this correlation function decays exponentially with the distance in the bulk, as it should be, since single-particle excitations have a spectral gap (in case of fermions), however, we see the recovery at the opposite edge. Hence, the edge modes, responsible for the similar recovery for the case $\mu_1=\mu_2$ and $\phi=0$ \cite{Dalmonte} are stable with respect to chemical potential difference and weak time-reversal symmetry breaking when processes breaking relative parity of the two sub-bands are neglected.

\begin{figure}
\includegraphics[width=8.0cm]{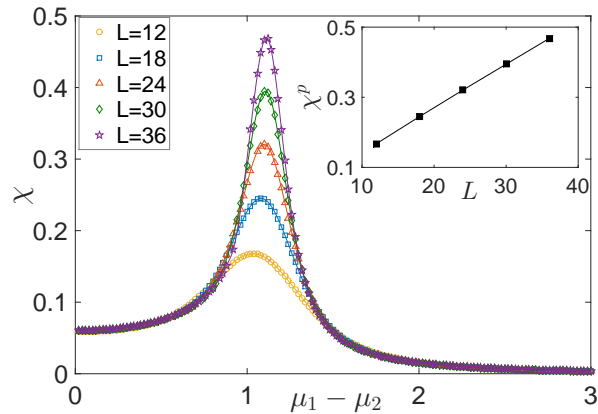}
\caption{ Behavior of fidelity susceptibility across the quantum phase transition from the two-channel to single-channel regime for $\phi=0$ when changing the relative chemical potential. Inset shows behavior of the peak value of fidelity susceptibility $\chi^P$ with the system size. For computational reasons in this particular simulation instead of fermions we use hard-core bosons and assume periodic boundary conditions. Density-density interactions are $g_{x}=1$, $g_{l,l'}\ll g_x$ corresponding to the case of extremely strong screening and pair-hopping amplitude is $W=1.8$ in Eq. (\ref{eq:hamiltonian}). Total particle density is kept constant, so that the filling corresponds to one-third particles in average per ladder site.}
\label{fig:FS}
\end{figure}
To study numerically the nature of the phase transition between the Tq1D state and single-channel regimes, we analyze the behavior of the ground-state fidelity susceptibility \cite{VenutiZanardi07,You+07,Gu10rev,us}. This is particularly convenient in our case, since we do not need to specify order parameter.
We study the (reduced) ground-state fidelity susceptibility with respect to the chemical potential difference,
\begin{equation}
\chi=-\frac{2}{L}\lim_{\delta \to 0} \frac{ \ln {| \langle   \psi_0 (\mu_1-\mu_2) | \psi_0 (\mu_1-\mu_2+\delta)  \rangle|}}{\delta^2}.
\end{equation}
The behavior of the fidelity susceptibility across the quantum phase transition, obtained from the minimal lattice model Eq. (\ref{eq:hamiltonian}) with unbroken time-reversal symmetry, for  $W=1.8$, $g_x=1$ and $g_{l,l'}\ll g_x$ is presented in Fig. \ref{fig:FS}. With increasing the system size the peak value of the fidelity susceptibility $\chi^P$ scales linearly, suggesting the Ising-type nature of the quantum phase transition, consistent with the bosonization analyses \cite{Rosch}. In fact, the phase transition predicted theoretically \cite{Rosch} is not of purely Ising nature, however, fidelity susceptibility studies presented here can not distinguish such details \cite{comment}. Also, it has been suggested that in the weak coupling limit, the phase transition is of $z=2$ character, similar to the Lifshitz point of non-interacting fermions, and only in strong coupling regime the effective Lorentz invariance emerges, producing $c=3/2$ criticality that is similar to Ising criticality superimposed on top of the overall gapless Luttinger-liquid charge mode. Our numerical simulations are done in strong coupling regime, where we resolve well-defined peak in fidelity susceptibility. To address the situation in the weak-coupling regime, we should consider both pair-tunneling as well as density-density type interactions much weaker than Fermi energy. For the moment, we can not comment on the nature of criticality in the weak coupling regime with the method that we use.

\begin{figure}
\includegraphics[width=7.50cm]{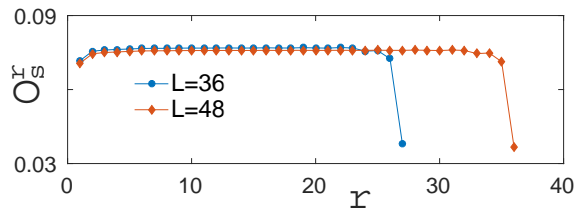}
\caption{Behavior of the string order defined in Eq. (\ref{stringorderfermions}) in Tq1D state as function of the distance $r$, starting from the rung $i=L/4$, indicating its long-range ordering in the bulk of the system. For string order, we present data for open boundary conditions, however, similar behavior (to that shown in the bulk) holds for systems with periodic boundary conditions as well. Simulation parameters are $\phi=g_{x}=g_{l,l'}=0$ and $W=1.8$ in Eq. (\ref{eq:hamiltonian}).}
\label{fig:Stringorder}
\end{figure}

In Fig. \ref{fig:Stringorder}, we plot the behavior of non-local string order parameter defined in Eq. (\ref{stringorderfermions}). We see that the string order saturates in the bulk, hence, the Tq1D superconducting state has indeed nonzero string order. 

\begin{figure}
\includegraphics[width=7.5cm]{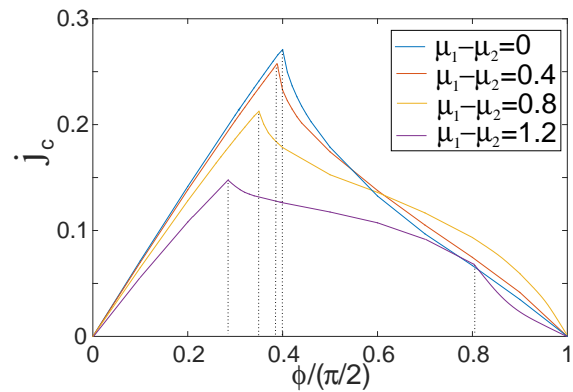} 
\caption{  Behavior of the chiral current as function of flux for different values of chemical potentials for $\phi=g_{x}=g_{l,l'}=0$ and $W=1.8$.}
\label{fig:ChC}
\end{figure}

A useful quantity for determining phase boundaries numerically in the presence of flux is the chiral current \cite{Pir,Gre} defined as
$j_c=\langle \frac{\partial H}{\partial \phi} \rangle$. In Fig. \ref{fig:ChC}, we plot the behavior of chiral current as function of the flux for different values of $\mu_1-\mu_2$. One can see that chiral current develops cusplike behavior at the phase boundaries, characteristic to C-IC phase transitions. In single-channel regime, the chiral current is strongly suppressed.

As already mentioned, the relative parity symmetry is only an effective, but not an exact symmetry of the microscopic model of quantum wire of spinless fermions Eq. (\ref{exact}), due to the assisted interband tunneling bulk processes and due to the boundary that provides particle scattering from one sub-band to the other.
We have checked that in the presence of time-reversal symmetry, the uniform assisted interband tunneling processes along the ladder do not lift the double degeneracy of the topological state and do not remove edge modes as long as they are weak and the ground state remains to be superconducting state. Value of the non-local string order parameter Eq. (\ref{stringorderfermions}) depends very weakly on amplitudes of the assisted interband tunneling processes.

We have also checked that a weak local single-particle tunneling (the most important process is tunneling at the edge, since in bulk single-particle states are gapped) also does not lift the double degeneracy of the ground state when time-reversal symmetry is present. In this respect the situation is similar to the case of exactly solvable fermionic ladder model realizing topological state \cite{Lang}. However, in the absence of time-reversal symmetry, we observe that when considering single-particle tunneling between the two legs at one of the edges only, the double degeneracy of the ground state is removed immediately (splitting between two ground states opens linearly with the amplitude of the local tunneling) and edge modes disappear. Interestingly, in the absence of time-reversal symmetry, when considering equal single-particle tunneling at both edges the two-fold degeneracy of the ground state is not lifted. Thus the important symmetry protecting the edge modes, when relative particle parity and time-reversal symmetry both are broken, is the reflection symmetry in the middle of the ladder that maps two edges onto each other. Hence, some discrete $Z_2$ symmetry is needed for protecting the topological state (relative particle parity or time-reversal or reflection symmetry in the middle of the ladder mapping two edges onto each other). In this respect our case looks similar to the symmetry protected topological Haldane state of spin-1 chain \cite{Oshikawa}.

\section{Interface between single-channel and Topological quasi-one-dimensional states}

\begin{figure}
\includegraphics[width=8cm]{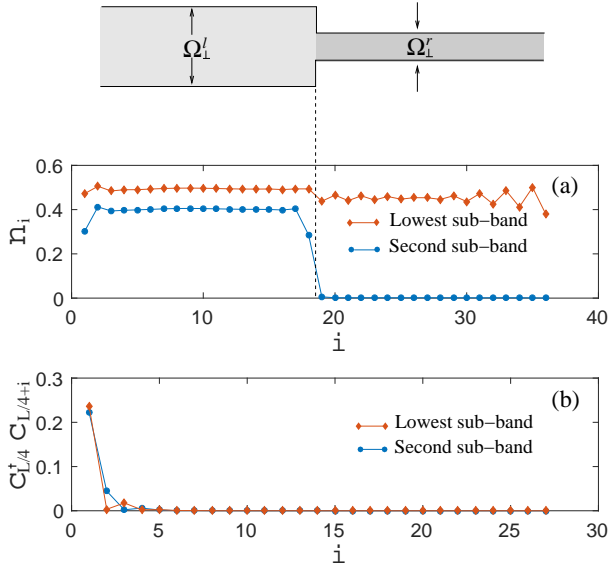}
\caption{    Interface between the two-channel and the single-channel regimes. The transversal lateral confinement frequencies are $\Omega^l_{\bot}<\Omega^r_{\bot}$. (a) Density profiles of electron in the lowest and second sub-bands for the chemical potential with a step-like behavior in the middle point of the system between the sites $L/2$ and $L/2+1$ (indicated by dashed vertical line) for $L=36$. The right part of the system is in the single-channel regime, with vanishing density in the second sub-band, and the left part is in the two-channel regime. (b) Single-particle correlation function from the bulk (here the middle point) of the two-channel regime towards the right direction shows exponential decay both in two-channel regime and across the interface. Similar exponential decay is observed towards the left side $\langle c^{\dagger}_{L/4}c_{L/4-j}\rangle\sim e^{-j\Delta}$. }
\label{fig:lead3}
\end{figure}

\begin{figure}
\includegraphics[width=9cm]{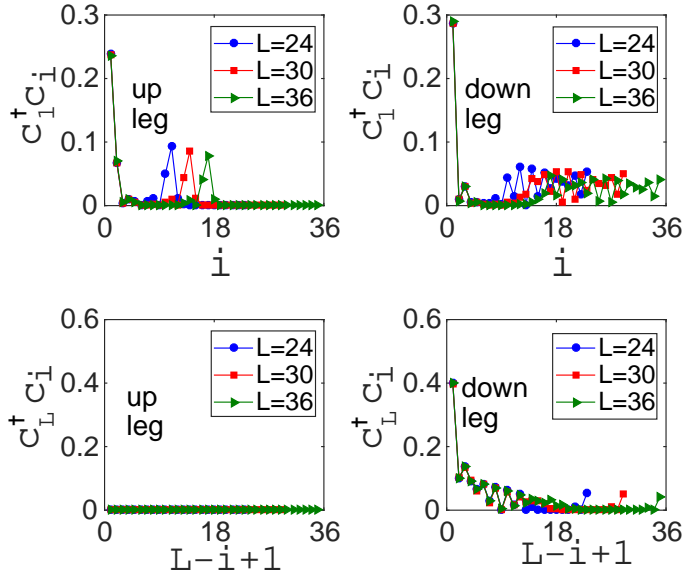}
\caption{Subband-resolved single-particle correlation functions from the edges corresponding to the case when an interface is present in the middle of the system, as indicated in Fig.\ref{fig:lead3}.}
\label{lead1}
\end{figure}

\begin{figure}
\includegraphics[width=7.5cm]{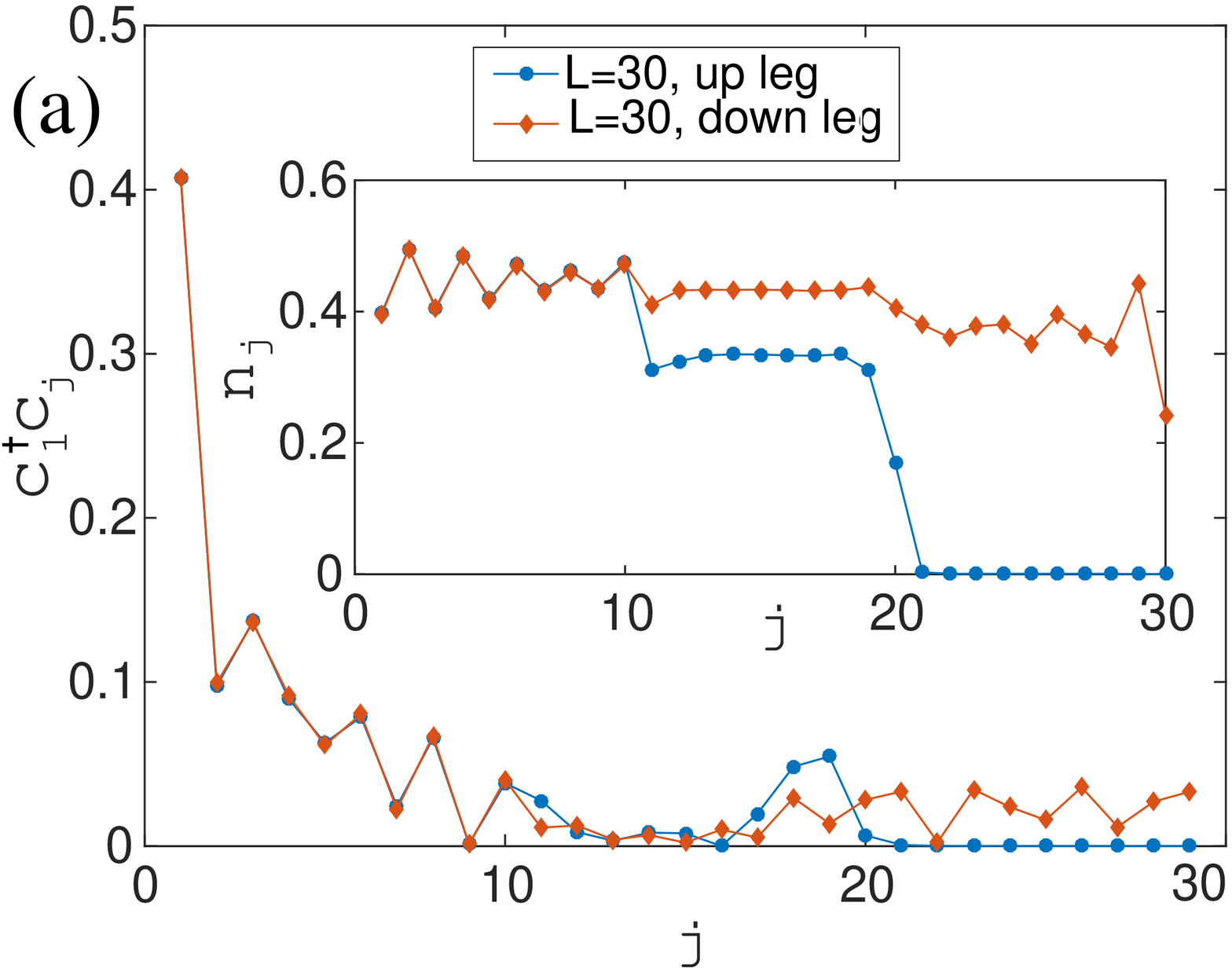}
\includegraphics[width=7.5cm]{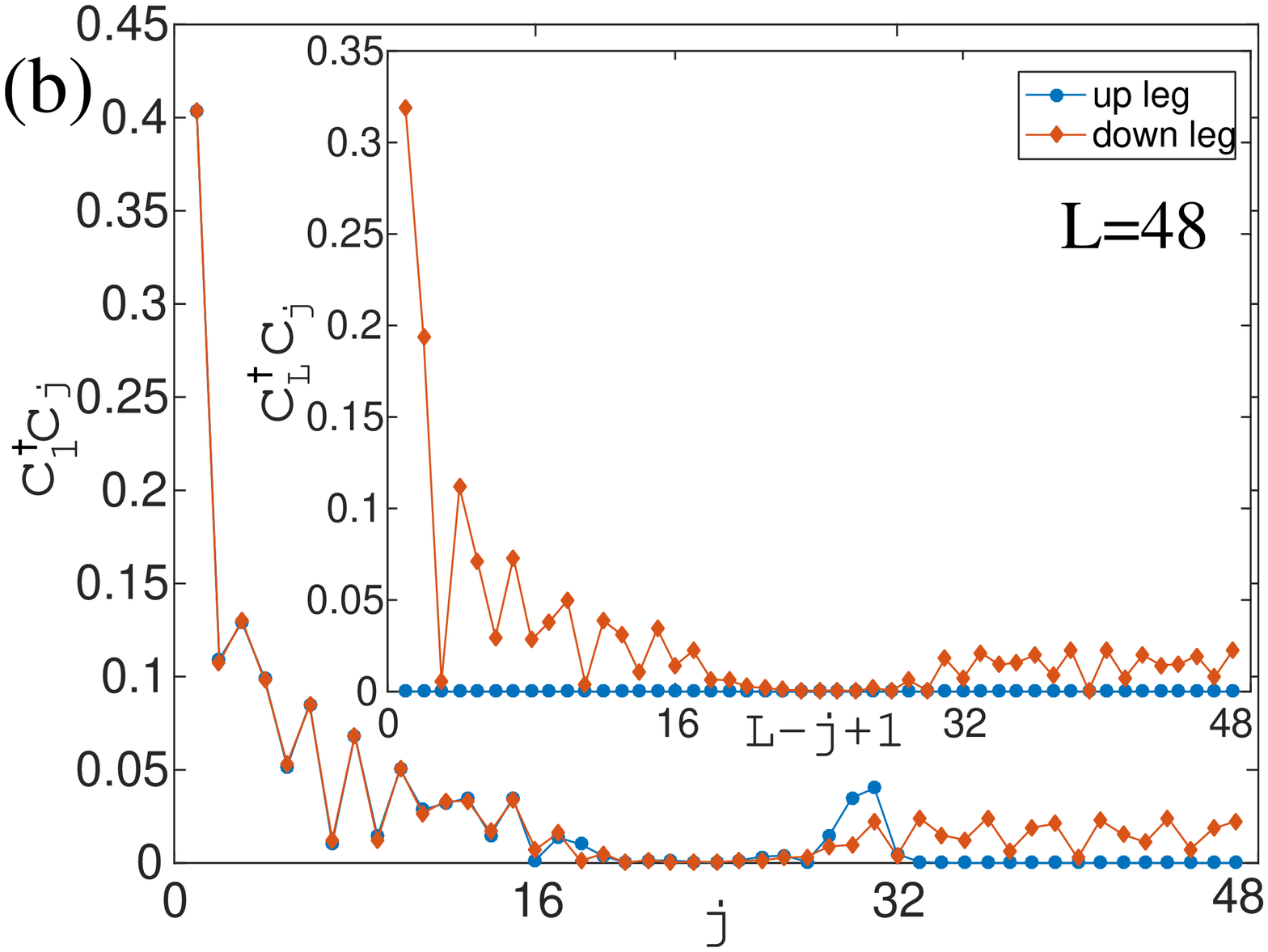}
\caption{Single-particle correlation functions in the presence of two interfaces: in the left side there is a two-channel state of non-interacting fermions, in the middle there is a Tq1D state and in the right side there is a single-channel state. Different system sizes are shown (a) L=30 rungs and (b)L=48 rungs. Inset in (a) shows sub-band resolved density profiles. }
\label{fig:LTT}
\end{figure}

\begin{figure}
\includegraphics[width=8.5cm]{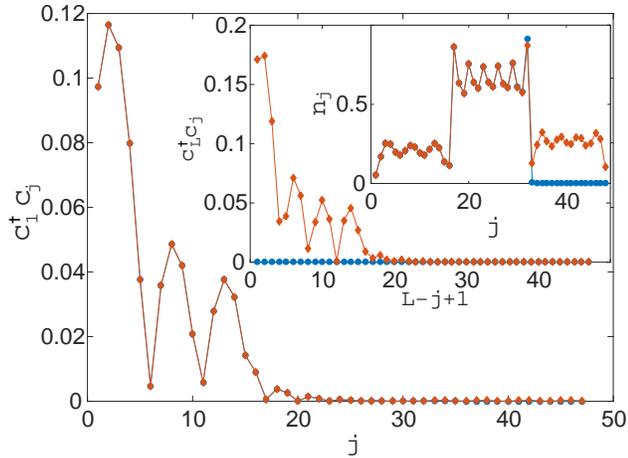}
\caption{Case of two interfaces as in Fig. \ref{fig:LTT}, however here in the middle part fermions interact attractively and densities in both legs are equal so that single-particle excitations have spectral gaps and there are no edge modes. Single particle correlation function from lead to single-channel regime decays exponentially, as opposed to algebraic recovery shown in Fig. \ref{fig:LTT}.
In the inset single-particle correlation function starting from the edge of the single-channel regime and density profiles in 3 different regions can be seen. Here $L=48$ rungs.}
\label{fig:lead5}
\end{figure}

In this section, we consider presence of an interface between the single-channel and topological quasi-one-dimensional states. Chemical potential changes as a step function at the interface so that densities have profiles as depicted in Fig. \ref{fig:lead3} (a). In  Fig. \ref{fig:lead3} (b) single-particle correlation functions are presented starting from the bulk of the Tq1d state and showing exponential decay to any other distant point. This is so, because in Tq1d state single-particle excitations are gapped. 

In all figures presented in this section the parameters for topological state are $\phi=g_{x}=g_{\ell,\ell'}=0$ and $W=1.8$. In Fig. \ref{lead1} we present correlation functions starting from the edge of the Tq1D state and extending all the way towards the another edge. As opposed to the behavior of correlation function starting from the bulk of the Tq1D state, shown in  Fig. \ref{fig:lead3} (b), the correlation function from the edge shows recovery over the interface, due to the edge modes.

Finally, we consider the case of two interfaces. System is made of a Fermi liquid lead, that is modeled by non-interacting electrons, a Tq1D intermediate state and a single-channel regime. This kind of setup may be relevant to describe realistic situations when lead is connected to a quantum wire and inbetween single-channel regime and the lead q1D state of electrons forms. Even though single particle correlation functions that start from lead decay exponentially in intermediate Tq1D region, they show recovery in single-channel regime as presented in Fig. \ref{fig:LTT}. When we place instead of Tq1D state CDW$_{\pi}$ (or other partially gapped) phase inbetween free fermions and single-channel regime we do not observe recovery of single-fermion correlations from lead to single-channel region. We present in Fig. \ref{fig:lead5} the case where Tq1D intermediate region is replaced by attractively interacting fermions with gapped relative excitation and gapless charge mode. One can see that single particle correlation function from lead to single-channel regime shows exponential decay and no recovery.

\section{Conclusion}

Using combination of effective field theory bosonization and DMRG simulations we have shown that when starting populating the second lowest transverse sub-band of interacting one-dimensional 
spinless electrons the superconducting quasi-one-dimensional state is stabilized that is a topological state characterized by non-local string order and hosting $Z_2$ discrete symmetry protected zero-energy edge modes for open boundary conditions. We have presented the single-particle correlation functions for the case when a spatial interface between the single-channel and Tq1D regimes is present, and showed that due to the edge modes of intermediate Tq1D state correlation fuction from lead to single channel regime shows algebraic recovery, even though that in the intermediate phase they are suppressed exponentially.

Based on our findings, we expect that upon increasing transversal confinement frequency, the quasi-one-dimensional zigzag state will smoothly evolve not into the superconducting state \cite{Meyer}, 
but into the CDW$_{\pi}$ state. The zigzag state can not be smoothly connected to the superconducting state because, as we showed, the nature of the superconducting state crucially depends on the statistics of the particles, whereas the nature of the zigzag state is independent of statistics. Moreover, zigzag state is characterized by the string order parameter \cite{Ruhman}, which is similar to the one found in CDW$_{\pi}$ state.

 This work has been supported by DFG Research Training Group (Graduiertenkolleg) 1729 and center for quantum engeneering and space-time research (QUEST). Part of the work of G.S. was done at the Institute of Theoretical Physics, Leibniz University of Hanover. We thank M. Baranov, H. P. B\"uchler and E. Jeckelmann for helpful discussions and J. Ruhman for comments on an earlier version of the manuscript. 



\begin{thebibliography}{10}
\bibitem{comment} Even though $c=3/2$ criticality was suggested to occur, effective velocity at critical point was argued to vanish in infrared limit. The finite-size scaling of the peak value of the fidelity susceptibility is sensitive to the dependence of correlation length on the deviation from criticality and not to the velocity of soft excitations at critical point. Hence vanishing effective velocity is not manifested in fidelity susceptibility studies and for this reason transition that we observe looks like criticality at ordinary quantum Ising tranisiton, with correlation length critical exponent equal to unity.
\end{thebibliography}
\end{document}